# A platform for cognitive monitoring of neurosurgical patients during hospitalization

Omer Ashmaig[1,3], Liberty S Hamilton [1,2,8], Pradeep Modur[5], Robert J Buchanan[1,5,6,7], Alison R Preston[2,3,4,9], Andrew J Watrous[1,2,3,4]

[1] Department of Neurology, Dell Medical School, The University of Texas at Austin, Austin, TX 78712
[2] Institute for Neuroscience, The University of Texas at Austin, Austin, TX 78712
[3] Center for Learning and Memory, The University of Texas at Austin, Austin, TX 78712
[4] Department of Psychology, The University of Texas at Austin, Austin, TX 78712
[5] Seton Brain and Spine Institute, Division of Neurosurgery, Austin, TX 78701
[6] Department of Neurosurgery, Dell Medical School, The University of Texas at Austin, 78712
[7] Department of Psychiatry and Behavioral Sciences, Dell Medical School, The University of Texas at Austin, 78712
8 Department of Speech, Language, and Hearing Sciences, Moody College of Communication, The University of Texas at Austin, 78712
[9] Department of Neuroscience, The University of Texas at Austin, Austin, TX 78712

Correspondence: andrew.watrous@austin.utexas.edu

**Abstract:** Intracranial recordings in epilepsy patients are increasingly utilized to gain insight into the electrophysiological mechanisms of human cognition. There are currently several practical limitations to conducting research with these patients, including patient and researcher availability and the cognitive abilities of patients, which limit the amount of task-related data that can be collected. Prior studies have synchronized clinical audio, video, and neural recordings to understand naturalistic behaviors, but these recordings are centered on the patient to understand their seizure semiology and thus do not capture and synchronize audiovisual stimuli from tasks. Here, we describe a platform for cognitive monitoring of neurosurgical patients during their hospitalization that benefits both patients and researchers alike. We provide the full specifications for this system and describe some example use cases in perception, memory, and sleep research. Our system opens up new avenues to collect more data per patient using real-world tasks, affording new possibilities to conduct longitudinal studies of the electrophysiological basis of human cognition under naturalistic conditions.

**Introduction**

There are currently 65 million active cases of epilepsy worldwide and approximately thirty percent of these patients are resistant to current medications (Kwan et al., 2011, Feigin et al., 2019). In these medication-resistant cases of epilepsy, patients may undergo invasive intracranial monitoring using indwelling electrodes to localize the source of their seizures. Depending on the purview of the clinical team and the individual case, patients may be implanted with multiple electrodes which record from hundreds of locations simultaneously. These patients often stay in the hospital for at least a week, presenting rare opportunities to directly measure local field potential (LFP) and/or single neuron activity in the behaving human brain over days or weeks with high spatiotemporal resolution (Jacobs & Kahana, 2010; Parvizi & Kastner 2018).

Given the insights which can be gained from direct electrophysiological recordings, it follows that maximizing the amount of data collected in this setting will prove beneficial to furthering our understanding of the human brain. However, there are several practical challenges when collecting human intracranial recordings. First, testing in this patient population often requires finding a "goldilocks" testing window in which the research team and patient are both available to conduct research. The invasive nature of electrode implantation often results in both physical and cognitive challenges that can affect a patient's ability to focus on performing experimental tasks. For example, pain medications may cause drowsiness, and the physical connections of electrodes make patient mobility a challenge. Thus, to allow for optimal testing conditions and prevent interruptions during testing, the patient must feel physically and cognitively well enough to perform cognitive tasks. Researchers must prioritize the needs of the patient and clinical team, so the research team frequently remains physically present until a testing window becomes available. These windows may occur during nights or weekends and place a burden on the research team to work beyond traditional work hours.

Second, researchers must employ tasks that accommodate a variety of cognitive abilities and impairments in epilepsy patients (Holmes 2015, Motamedi & Meador 2003). Practically, the researcher may ask an uncomfortable patient to perform a cognitive task with focus and effort instead of watching TV or browsing the internet. Humans prefer tasks which are appropriately challenging for their skillset (Csíkszentmihályi et al., 1990). Thus, if an experiment is too easy or too demanding, the patient may not agree to perform the task at all or may only perform it once, reducing the amount of data collected per patient. This can be even more problematic when multiple research groups are working with the same patient. As the patient performs different tasks, they naturally gravitate towards tasks which are engaging and are appropriately challenging. Researchers with less suitable tasks may therefore be unlikely to obtain more than one session of data. This can also be a challenge when more than one session of a task is required for adequate statistical power. All together, the practical limitations of finding the "goldilocks" testing window and employing an appropriate task reduce the amount of usable data that can be gained from each patient.

Despite these challenges, cognitive neuroscientific inquiry has increasingly leveraged human intracranial recordings (Jacobs et al., 2010; Parvizi & Kastner, 2018; Brazier 1968; Leonard and Chang 2016) . Born out of the traditions of cognitive psychology and stimulus-response views of cognition, most of this work has measured behavior using "classical" tasks with experimenter-generated stimuli. While such well specified and controlled designs benefit researchers, these designs often have seemingly arbitrary task demands from the patients point of view. Thus, while classical tasks provide tight experimental control, they may do so at the expense of ecological validity. The arbitrary and repetitive nature of such tasks may limit a patient's enthusiasm to repeat experiments, reducing the amount of data collected per patient. Given the value of human intracranial recordings to cognitive neuroscientific progress, research protocols which collect the maximum amount of useful behavioral and neural data will expedite our understanding of cognition.

It is increasingly recognized that neuroscientific models should be developed and tested in more real-world contexts (Nastase et al., 2020; Hamilton & Huth 2020; Matusz et al., 2019; Yoder & Belmonte 2010). For example, recent work has focused on understanding cognition using naturalistic stimuli during memory encoding (Baldassano et al., 2017; Chen et al., 2017; Davis et al., 2020; Heusser et al., 2020; Antony et al., 2021;Michelmann et al., BiorXiv) or spatial navigation (Stangl et al., 2020). While naturalistic experimental designs present additional challenges in understanding the data compared to classical paradigms, naturalistic tasks such as video games can provide novel benefits and insights into a diverse range of cognitive mechanisms (Boot 2015; Palaus et al., 2017). For example, video games have been shown to enhance motivation compared to traditional neuropsychological tasks (Lohse et al., 2013; Ferreira-Brito et al., 2019; Reid 2012). Furthermore, the real world does not operate in a uni- or bi-modal fashion as the brain must perceive and respond to multisensory information (Sella 2014; Atteveldt et al., 2014; Stein & Stanford 2008). More real-world experimental designs may therefore not only motivate patients to perform more tasks during their hospitalization but also may enhance behavioral performance and lead to unique cortical activity compared to "classic" experiments (David, Vinje, Gallant 2004; Bijanzadeh 2020; Matusz 2019; Sella 2014).

To address the above challenges, we describe here a platform for continuous cognitive monitoring of epilepsy patients undergoing intracranial monitoring during their hospitalization (Figure 1, Figure 2). This platform aims to maximize the amount of useful behavioral and neural data per patient and further our understanding of the working mechanisms of cognition during naturalistic behaviors. This system can provide novel insights by adding content and external validity to the current "classic" neuroscientific literature.

**Materials and equipment**

- PlayStation 4 with controller
- Raspberry Pi
- HDMI to DVI converter
- DVI to HDMI converter
- HDMI splitter
- Hauppauge Standalone Video Recorder
- Micro USB cable (8 foot minimum)
- Monitor or TV with HDMI port
- USB A or mini USB A cable (depends on StimTracker model)
- Cedrus StimTracker Duo or Quad
- 3 HDMI to HDMI Cables
- Mobile AV Cart
- Power Extension Cord
- Large Hard Drive (at least 1 TB)
- Python and Bash Scripts (supplemental)

**Methods**

Our testing platform synchronizes continuous neural recordings with audiovisual stimuli and button-press responses (Supplemental Movie 1), allowing researchers to perform analyses between different recording modalities. This system allows researchers to synchronize continuous neural recordings with any activity on the PlayStation 4 console. All system components are placed on a mobile cart that is rolled into the patient's room and placed near the foot of the bed so that the monitor can swing out over the patient's feet. Each button press is logged, overlays a unique auditory tone within the audiovisual recording, and sends a customizable, jitter free event marker, which are delayed in the neural recordings by precisely 2 ms. This setup allows for precise offline syncing between the neural recordings, audiovisual stimuli, and motor responses. Through the PlayStation 4 console, researchers have a spectrum of possibilities for tasks - from using currently available video games and movies to fully designing a well-controlled experiment through a game creation system such as

"Dreams", a video game that allows users and/or researchers to create their own games (Skrebels 2020).

In our implementation, a stand-alone video recorder continuously records audiovisual media from the gaming console to an external hard drive. The PS4 controller sends button press events to the Cedrus StimTracker via a Raspberry Pi, allowing for offline synchronization of behavioral and neural data. The Raspberry Pi runs two python scripts simultaneously. A Python 2 script detects and logs controller activity via a micro USB cable and sends an event code to a Cedrus StimTracker via a USB cable. Each button corresponds to a unique, prime-numbered TTL pulse width embedded in the neural recordings. A Python 3 script saves a unique tone into the audiovisual recording for each button press and patients do not hear these tones. These scripts are available at the following Github page: github.com/oeashmaig/CCM.

For our particular configuration, an HDMI splitter from the PS4 system to the audiovisual recorder resolves compatibility issues between the two devices and splits the video signal to a second external hard drive and monitor, enabling family and friends to watch alongside the patient. A second hard drive of the original audiovisual signal becomes useful here to remove the button tones during offline processing.

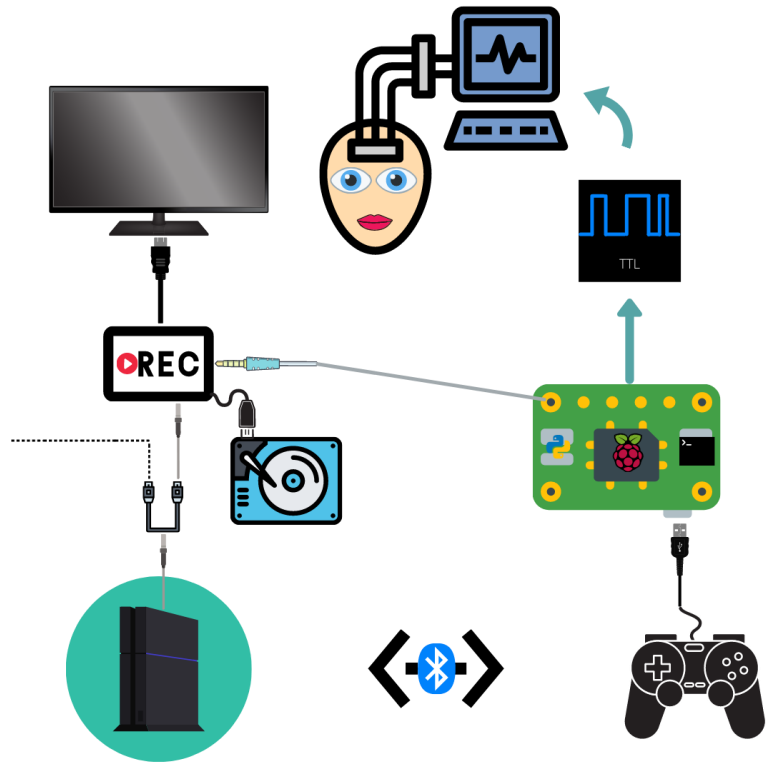

**Figure 1: Schematic depicting the connections between each component.** The PS4 system outputs audiovisual stimuli to a video recorder through an HDMI splitter. The video recorder saves the audiovisual media to an external hard drive and then outputs the signal to a monitor. The Raspberry Pi detects and logs PS4 controller button responses and sends a unique tone for each button which is embedded in the audiovisual recording. The Raspberry Pi simultaneously sends event markers to the Cedrus StimTracker, which generates a unique TTL event recorded alongside the neural data.

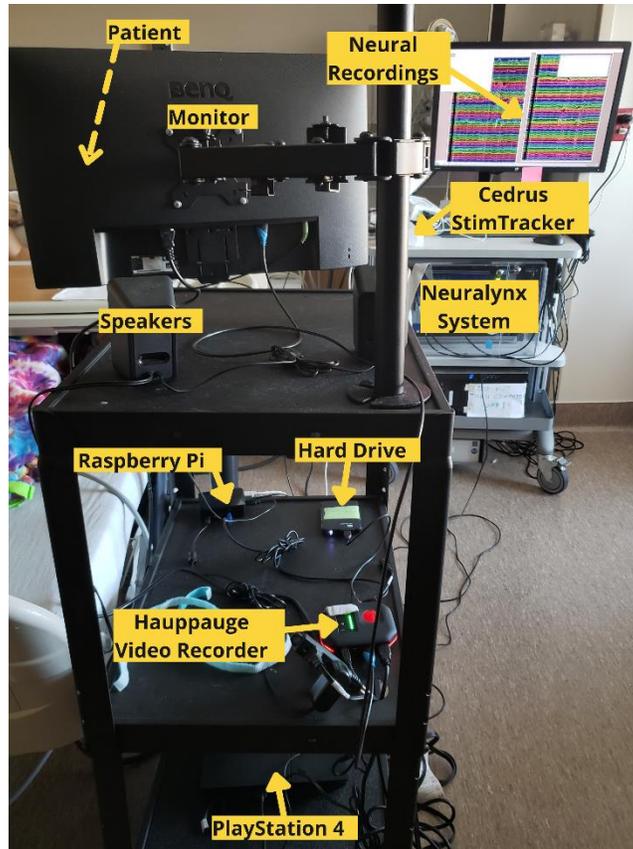

**Figure 2: Labeled photograph of the cognitive monitoring system in use in the hospital setting.**

Step-by-step procedures:
Basic instructions are given here. Full line-by-line details on installation are also provided on the github page (http://github.com/oeashmaig/CCM).

*Setting up the Raspberry Pi*

1. Download the python and bash scripts from Github
2. Install the required python 2 and 3 modules: qt5-default, libasound2-dev, pyxid, simpleaudio, pygame
3. Download and install the D2xx driver from FTDI chip (https://ftdichip.com/drivers/d2xx-drivers/)
4. Edit the /etc/rc.local file to run the bash script after startup

*Initial setup of the testing platform*

1. Connect the PS4 controller to the PS4 system via Bluetooth before booting the Raspberry Pi
2. On the PS4 system, navigate to "System Settings" and disable HDCP
3. Plug in the HDMI-to-DVI-to HDMI converter to the PS4 system
4. Connect the HDMI output from the PS4 to an HDMI splitter to the Hauppauge Video Recorder, which is connected to a storage device
5. Configure the Hauppauge device and connect it to a high capacity external hard drive (format as Windows NTFS with Master Boot Record , ≥ 2TB recommended)

6. Connect an audio cable from the Raspberry Pi to the Hauppauge Video Recorder
7. Connect HDMI cable from Hauppauge Video Recorder to TV/monitor
8. Connect the controller to the Raspberry Pi via a long micro USB cable
9. Connect the Raspberry Pi to the Cedrus StimTracker via a USB A cable

*Utilizing the system with neural recordings for patient testing*

1. Begin neural recordings using NeuraLynx research electrophysiology system
2. Ensure all system components are plugged in and all devices have power
3. Turn on the PS4
4. Ensure the Hauppauge shows a green LED and then start recording (indicated by a solid red LED).
5. Instruct the patient on how to use the system, including streaming services, various games tasks, and other downloadable multimedia.
6. Confirm button presses are being detected as event markers in the NeuraLynx system.

**Anticipated Results**

To date, we have deployed the cognitive monitoring system in 4 patients. Each patient has used the system for several hours to play a variety of movies, TV shows, and games during simultaneous neural recordings. All patients watched the first episode of Sherlock along with patient-specific movies and games based on their preferences. These data will enable investigation of neural correlates of volitional behavior for patient-specific choices. In addition, future analyses using these data will focus on generating spatiotemporal receptive fields using the movie data and spectrotemporal receptive fields using the audio data. However, rather than focusing on a specific research question, the focus of this manuscript is to describe the system and highlight its possible uses. Below, we describe these anticipated advantages and potential disadvantages.

*Advantages*
Researcher benefits

The primary benefit to researchers is the ability to collect more neural and behavioral data per patient while providing a common set of tasks that all patients can perform regardless of cognitive ability. Our system provides a means for patients to entertain themselves while providing useful neural recordings with millisecond-precision, synchronized stimuli at their leisure. For example, we have found our system to be particularly useful in patients undergoing sleep deprivation to induce seizures. These patients are often looking for something to do to stay awake late at night and thus readily engage in low-effort cognitive tasks such as playing video games or watching movies. This platform thus reduces the need for the research team to be physically present in the hospital, mitigating issues with patient and research team availability.

Our system also affords the opportunity for researchers to capture activities of daily living, such as listening to music, watching streaming audio or video services, or playing video games. This allows researchers to investigate a diverse array of cognitive functions within a naturalistic context. Even in periods of fatigue, most patients have been quite willing to watch Netflix "for science". We anticipate that this increase in multimodal data will allow researchers to build more robust within-patient models of cognitive processes.

Patient benefits

Many patients spend most of their time in a hospital bed for at least a week and become quite bored. Our platform provides the patient with an alternative to the hospital-provided television, their primary form of entertainment. At the same time, we leverage each patient's unique interests to investigate a wide range of cognitive processes. The patient has the choice to either watch their favorite shows or play a pre-selected library of games. Naturally, these two options are less cognitively demanding for the patient compared to "classic" tasks and minimizes subjective feelings associated with volunteering time for a typical task. Furthermore, our typical patient population has a variable and limited number of controlled experiments they can perform in a single day without significantly losing motivation and energy. Thus, patients can provide additional useful data on their own schedule based on their own preferences and choices. Finally, the testing platform reduces possible experimental bias and/or stress experienced by patients due to being observed by a researcher (Yoder & Belmonte 2010).

Video game controllers are designed to be a more intuitive interface for patients compared to using arbitrary keyboard mappings or a buttonbox, minimizing the need for patient training. All patients thus far have either had some familiarity with using the PS4 or required no more than 5 minutes of instruction. We also provide a "cheat sheet" for patients to understand what they could do on our system, how to do it, and a simple explanation of why it would improve our understanding of the human brain.

Additional Potential Use Cases

We anticipate that our system can benefit several other research endeavors. First, video games induce seizures in a small subset of epileptic patients. In cases of reflex and musicogenic epilepsy, some patients experience seizures due to photic or auditory stimulation (Stern 2015; Ferrie et al 1994; Millet et al 1999). However, the underlying mechanisms of seizure propagation in these patient populations remains unclear. Furthermore, in non-photosensitive epilepsy patients, the current literature is unclear whether video games induce seizures due to non-photic factors such as changes in arousal or simply due to chance (Ferrie et al 1994; Millet et al 1999). Through continuous cognitive and electrophysiological monitoring preceding seizures, both clinicians and researchers may better understand the underlying pathophysiology and manifestation of seizures on both an individual and population level.

Second, there has been a growing interest in understanding the role of sleep in systems memory consolidation, especially the role of sharp wave ripples (Jiang et al., 2019). These studies have been primarily studied in rodents given the rare and limited ability to investigate these processes directly in human recordings. Thus, the use of our testing platform presents the opportunity to significantly increase the amount of recorded human neural data during sleep. Finally, our system employs a PS4 system to collect neuroscientific data which could also be distributed to the general public to collect a large sample of normative behavioral data.

Flexibility of research recording setup

While we have presented a specific use case here for a NeuraLynx system with a Cedrus StimTracker, our setup is also transferrable to other research recording systems. For example, with Tucker Davis Technologies (TDT) RZ2 hardware including analog and digital inputs, the Cedrus StimTracker can be bypassed altogether, and audio button press signals may be recorded directly from the Raspberry Pi to an analog input that is synchronized with neural data. A similar process may be performed for split output audio signals such that one output goes to a speaker that the patient hears, and the other is passed as an analog input to the research system. We have tested this system successfully with TDT and NeuraLynx systems, but it should in principle work for any research system that is able to take in audio as an analog input or generate TTL pulses from the data as is shown here.

***Limitations and Potential Pitfalls***

There are downsides to implementing our system, although we consider these minor compared to its advantages. First, more equipment and setup time is needed, though our typical setup time for the system is under 15 minutes.  Second, the additional electronic components in the patient's room could introduce 60Hz noise to the neural recordings. Notably, we have not found this in our recordings to date and this issue could be further mitigated with an uninterruptible power supply. Given the significant increases in recorded behavioral and neural data, more storage is required on both data acquisition systems. Typically, neural recordings require approximately 2 TB of storage per patient and the audiovisual recordings require 100 GB of storage per day.

This significant addition of behavioral and neural data requires additional computational resources during offline processing. Although more data may be obtained from video games in some patients, most current video games are considered to be less statistically powerful per unit time for a specific hypothesis and may present challenges in removing or isolating confounding stimulus features (Hamilton & Huth 2020). However, at the same time, we can also obtain more behavioral responses per unit time, especially in motor-related tasks. In addition, the variety of acoustic and linguistic information present in hours of Netflix movies may surpass the variety that can be presented in a short "classic" task. In either case, recent computational advances in computer vision using deep learning models (Mathis et al. 2018) can help circumvent these issues (Hamilton & Huth 2020).

As previously discussed, patients may be intrinsically motivated to use this testing platform. This may introduce a potential issue for the research team when a patient prefers to use this platform over more "classic" experiments, although we have not yet observed this in our patient population. To mitigate this issue, we introduce our system to patients after we have collected data with our "classical" tasks.

In cases where a research team is interested in investigating various possible cognitive mechanisms through movies or TV, but is not physically present to observe the patient, an important question arises - how do you know if a patient is paying attention or if they are watching with loud TV volume? While we ask patients to pause during periods of conversation or other moments of distraction, clinical video recordings of the patient could be used to verify attention towards the screen. Using such recordings requires the researcher to follow the appropriate ethical guidelines and consent process. Eye tracking could also be added in future upgrades to our platform in order to more directly measure viewing behavior.

While the system is designed to give patients freedom of choice, researchers must narrow the list of possible media to be included on the system. If each patient watches a different movie based on their own preference, how would researchers perform group level analysis? We have addressed this issue by taking a hybrid approach and ask patients to watch a small subset of content on the system before watching whatever they choose. This should allow researchers to study how volitional behaviors and their associated neural dynamics augment task performance, for instance during memory encoding (Estefan et al., 2021; Fried et al., 2017; Gureckis & Markant 2012; Markant & Gureckis 2014; Voss et al., 2010). Currently, we ask patients to first watch

Sherlock and play Pac-Man before doing other tasks on the system, as these stimuli have been used in several previous fMRI studies (Chen et al., 2017; Baldassano et al., 2017; Vodrahalli et al., 2018). In cases where patients watch different shows, we can study language at a population level by using transcripts and computational models to find overlap in natural language representations. For auditory and visual tasks, the spectrogram of the audio or a wavelet decomposition of the visual information could be used to predict neural responses. Future studies are needed to determine the degree of generalizability of receptive fields fit on non-overlapping stimulus sets, however, our prior work has shown that this may be possible (Desai et al. Biorxiv 2021).

**Discussion**

We have provided the design, specifications, and code for implementing a continuous cognitive monitoring platform. We implemented this system for use with epilepsy patients undergoing invasive monitoring for seizures but believe it could be modified for standard EEG monitoring or other patient populations. Furthermore, while this platform is designed for use with a PlayStation 4 console and a Cedrus StimTracker, the general methodological approach can be easily modified with alternative hardware. As technology continues to innovate, we believe the idea of continuous cognitive monitoring is more important than the specific implementation of the system as described here.

Another consideration when developing methods to maximize data acquisition is the cognitive abilities of patients. Although not every patient can perform cognitively demanding tasks, nearly all patients can either play simple arcade-style video games or watch a movie. Movies and TV shows present opportunities to investigate event-segmentation, language, and memory processes within an ecologically-valid context, thus adding content validity to our understandings of diverse cognitive functions (Berezutskaya et al., 2020). We believe that expanding our neuroscientific research protocols to more complex, real-world contexts will provide novel insights into the working mechanisms of cognition.

Another issue likely to arise with our system is how researchers will make sense of increasing amounts of multi-modal data. Clearly, more computational resources will be required, and we anticipate that recent advances in statistical and computational models will allow us to analyze these complex data. For example, recent work using clinical audio and video recordings of patients have been used to characterize naturalistic behavior (Wang et al., 2016; Bijanzadeh et al., *BiorXiv*). Our system builds upon this work and provides a means to track what the patient is doing more closely than can be derived from clinical recordings alone. Moreover, the platform we describe is useful in gently guiding patients into performing particular cognitive operations, such as watching, listening, and playing. Furthermore, innovations in computer vision using deep learning models, such as DeepLab Cut (Mathis et al. 2018) require minimal training and should be able to accurately track task performance in individuals. Unsupervised learning models and dimensionality reduction methods can help us develop unbiased behavioral and neural insights into these multimodal recordings without *a priori* hypotheses or pre-defined features of interest (Hamilton & Huth 2018, Cabañero-Gómez et al. 2018; Wang et al., 2016). While many video games may be considered to be less statistically powerful (Matusz et al., 2019), video games have a greater natural effect size, meaning they add external and content validity alongside the detection and importance of an effect.

The system we describe can accommodate the spectrum of possible experimental designs from more "classical" to more naturalistic. "Classical" experiments are well-controlled and offer valuable insights, but lack applicability to real-world situations. In our view, the goal is not to get rid of "classical" experimental paradigms, but instead to present a method to collect additional and novel data, integrate both experimental designs into our existing workflow, and to enrich real-world neuroscientific research. While continuous cognitive monitoring is more complex and may be more challenging to analyze, we anticipate that it will offer content and external validity to diverse scientific fields in cognitive neuroscience. By expanding the analysis of neural recordings from well-controlled, simplistic paradigms to more complex stimuli, we believe our system can provide data to validate and complement the current neuroscientific literature.

**Supplemental Video 1**: Example synchronized behavioral and neural data generated using our testing platform. Upper panel shows PacMan gameplay. Simultaneously recorded neural data from an example hippocampal electrode and button press events are shown in the middle and lower panels, respectively.